\documentclass[twocolumn,showpacs,amsfonts,aps,prc,floatfix,%
superscriptaddress,nofootinbib]{revtex4-1} 

\usepackage{amsmath}
\usepackage{bm}
\usepackage{graphicx}
\usepackage{url}
\usepackage{longtable}

\newcommand{\be}[1]{\begin{equation}\label{#1}}
\newcommand{\ee}{\end{equation}}

\usepackage{color}

\begin{document}


\title{
Elliptic flow in Pb+Pb collisions at
$\sqrt{s_{NN}}$ = 2.76 TeV:\\
hybrid model assessment of the first data
}
\date{\today}

\author{Tetsufumi Hirano}
\email{hirano@phys.s.u-tokyo.ac.jp}
\affiliation{Department of Physics, The University of Tokyo,
Tokyo 113-0033, Japan}
\affiliation{Nuclear Science Division, Lawrence Berkeley National Laboratory,
Berkeley, CA 94720, USA}

\author{Pasi Huovinen}
\email{huovinen@th.physik.uni-frankfurt.de}
\affiliation{Institut f\"ur Theoretische Physik, Johann Wolfgang Goethe-Universit\"at, 60438 Frankfurt am Main, Germany  
}

\author{Yasushi Nara}
\email{nara@aiu.ac.jp}
\affiliation{Akita International University, Yuwa, Akita-city 010-1292, Japan}

\begin{abstract}
We analyze the elliptic flow parameter $v_{2}$
in Pb+Pb collisions at
$\sqrt{s_{NN}}$ = 2.76 TeV
and in Au+Au collisions at $\sqrt{s_{NN}}=200$ GeV
using a hybrid model in which the evolution of the quark gluon
plasma is described by ideal hydrodynamics with a state-of-the-art
lattice QCD equation of state, and the subsequent hadronic stage by
a hadron cascade model.
For initial conditions,
we employ Monte-Carlo versions of 
the Glauber and the Kharzeev-Levin-Nardi models
and compare results with each other.
We demonstrate that
the differential elliptic flow $v_2(p_T)$ hardly changes when the
  collision energy increases, whereas the
integrated $v_2$ increases
due to the enhancement of mean transverse momentum.
The amount of increase of both $v_2$ and mean $p_T$ depends
significantly on the model of initialization.
\end{abstract}

\maketitle


The recently started heavy ion program at Large Hadron Collider 
(LHC) in CERN
opens up opportunities
to explore the deconfined matter, the quark gluon plasma (QGP),
in a wider temperature region.
Elliptic flow \cite{Ollitrault},
which played an essential role
to establish the new paradigm of 
the strongly coupled QGP \cite{Gyulassy:2004vg,sQGP}
at Relativistic Heavy Ion Collider (RHIC)
in Brookhaven National Laboratory (BNL) \cite{BNL},
is one of the key observables at LHC
to investigate the bulk and transport properties of the QGP.
First elliptic flow data in Pb+Pb collisions
at $\sqrt{s_{NN}}$ = 2.76 TeV were recently published 
by the ALICE Collaboration \cite{Aamodt:2010pa}.
The first goal of flow measurements is to see
whether hydrodynamic models reproduce the flow as well at LHC as at
RHIC, and thus whether the QGP depicts similar strong coupling
nature at LHC.


This Rapid Communication is a sequel to our previous
work~\cite{Hirano:2010jg} where we predicted the elliptic flow
parameter $v_2$ before any LHC data was available. In this
publication we take the advantage of the first LHC
data~\cite{Aamodt:2010pa,Aamodt:2010pb} to fix the final particle
multiplicity, which removes the main uncertainty in our prediction,
and allows us to use a Glauber type initialization too. We calculate
the elliptic flow parameter $v_{2}$ and its transverse momentum
($p_{T}$) dependence in Pb+Pb collisions at LHC and compare them
with the data. Our model for the space-time evolution of the matter
is the same we used in Ref.~\cite{Hirano:2010jg}: A hybrid model
where the expansion of the QGP is described by ideal
hydrodynamics~\cite{Hirano:2001eu},
and the subsequent evolution of hadronic matter below switching
temperature $T_{\mathrm{sw}} =155$ MeV, is described using a hadronic
cascade model JAM~\cite{JAM}.
During the fluid dynamical stage, we
employ EoS $s95p$-v1.1, 
which interpolates between
hadron resonance gas at low temperatures and recent lattice
QCD results by the hotQCD collaboration~\cite{Cheng:2007jq,Bazavov:2009zn}
at high temperatures
in the same way as $s95p$-v1 \cite{Huovinen:2009yb},
 but the hadron resonance gas part 
contains the same hadrons and resonances as the JAM hadron
cascade~\cite{JAM}. The details of the interpolating procedure are
explained in Ref.~\cite{Huovinen:2009yb} and the parametrization and EoS
tables are available at Ref.~\cite{EoSsite}.

Initial time of hydrodynamic simulations
 is fixed to be $\tau_{0}$ = 0.6 fm/$c$
throughout this work.
For initial conditions in the longitudinal direction,
we assume the Bjorken scaling solution \cite{Bjorken:1982qr}.
To initialize the density distributions in the transverse plane,
we utilize two Monte-Carlo
approaches: Monte-Carlo Glauber (MC-Glauber) model
\cite{Miller:2007ri} 
and Monte-Carlo Kharzeev-Levin-Nardi (MC-KLN) model~\cite{MCKLN}.
Using these Monte-Carlo models,
we calculated initial conditions for hydrodynamic 
simulations in the transverse plane
with respect to the \textit{participant plane}
in our previous work \cite{Hirano:2010jg}.
These initial density profiles contain
effects of eccentricity fluctuation on average.
However, the ALICE Collaboration mainly obtained $v_2$ 
using the 4-particle cumulant method $v_{2}\{4\}$~\cite{Borghini:2000sa}, 
which is expected to contain less eccentricity fluctuation 
and non-flow effects than $v_{2}$ measured 
using the 2-particle cumulant method,
$v_{2}\{2\}$\footnote{
If the eccentricity
distributes exactly as Gaussian 
event-by-event, effects of 
eccentricity fluctuation vanish
in $v_{2}\{4\}$ \cite{Voloshin:2007pc,Ollitrault:2009ie}.}.
Therefore
we calculate in this Rapid Communication initial profiles
with respect to the \textit{reaction plane}:
We average over many events using Monte-Carlo calculations
instead of shifting and rotating a distribution
event-by-event
to match the main and sub axes of the ellipsoids
as was done in the previous work \cite{Hirano:2010jg,Hirano:2009ah}.
It should be noted that
the distributions obtained in this way
are not identical to the ones
from the optical Glauber model or the factorized KLN (fKLN) model \cite{fKLN}
due to finite nucleon size effects \cite{MCKLN}: 
the collision points 
in the transverse plane 
are smeared using inelastic cross section of $p+p$ collisions
in the ``mean-field'' option in the Monte-Carlo approach \cite{MCKLN}
to obtain smooth initial conditions
for hydrodynamic simulations.

In the MC-KLN model, we calculate 
distribution of gluons at each transverse grid
using the $k_t$-factorized formula~\cite{KLN}.
Using the thickness function $T_{A}$,
we parametrize the saturation scale for a nucleus $A$ as
\begin{equation}
Q_{s,A}^2 (x; \bm{x}_\perp)  =  2\ \text{GeV}^2
\frac{T_A(\bm{x}_\perp)}{1.53\ \text{fm}^{-2}}
\left(\frac{0.01}{x}\right)^{\lambda}
\label{eq:qs2}
\end{equation}
and similarly for a nucleus $B$.
We choose $\lambda=0.28$ and a proportionality constant
in the unintegrated gluon distribution in the $k_{t}$-factorized formula
to reproduce centrality dependence of $p_{T}$ spectra 
obtained by the PHENIX Collaboration \cite{Adler:2003cb}.
As a default parameter set at LHC,
we use the same parameters except for colliding energy
and mass number of incident nuclei.
This predicted $dN_{\mathrm{ch}}/d\eta \sim 1600$
at 5\% most central collisions \cite{Hirano:2010jg}, which 
turns out to be consistent
with the recent ALICE measurement \cite{Aamodt:2010pb,ALICEdNdeta}.

In the MC-Glauber model,
one calculates the number distributions of participants 
$\rho_{\mathrm{part}}$ and
of binary collisions $\rho_{\mathrm{coll}}$
for a given nuclear density distribution.
We model the initial entropy distribution
in hydrodynamic simulations
as a linear combination of
$\rho_{\mathrm{part}}$ and  
$\rho_{\mathrm{coll}}$ in the transverse plane:
\begin{eqnarray}
\frac{dS}{d^{2}\bm{x}_{\perp}} & \propto &
\frac{1-\alpha}{2}\rho_{\mathrm{part}}(\bm{x}_{\perp})
+ \alpha \rho_{\mathrm{coll}}(\bm{x}_{\perp}).
\label{eq:dsdx2}
\end{eqnarray}
At the RHIC energy, the mixing parameter $\alpha = 0.18$
and the proportionality constant in Eq.~(\ref{eq:dsdx2})
are chosen to reproduce the centrality
dependence of $p_{T}$ spectra
at RHIC \cite{Adler:2003cb}. 
We tune these two parameters 
in Pb+Pb collisions at LHC 
to reproduce
the centrality dependence of charged hadron multiplicity \cite{ALICEdNdeta}.
For both initializations we do the centrality
cuts according to the $N_{\mathrm{part}}$ distribution
from the MC-Glauber model \cite{Hirano:2010jg}

\begin{figure}[htb]
\includegraphics[width=3.4in]{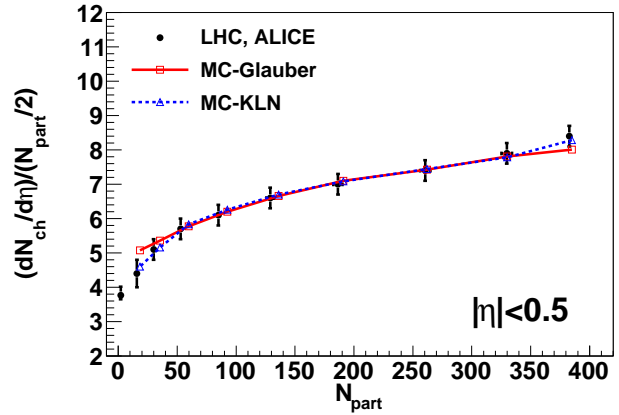}
\caption{(Color online)
Centrality dependence of charged hadron multiplicity
in the MC-Glauber and the MC-KLN initialization
is compared with ALICE data \cite{ALICEdNdeta,Aamodt:2010ft}.
A data point from inelastic events
at $\sqrt{s_{NN}}=2.36$ TeV in $p+p$ collisions \cite{Aamodt:2010ft}
is shown at $N_{\mathrm{part}}=2$.
Each point in theoretical results
from right to left corresponds to 0-5, 5-10, 10-20,
20-30, 30-40, 40-50, 50-60, 60-70, and 70-80\% centrality, respectively. 
}
\label{fig:nchnpart}
\end{figure}

\begin{figure}[htb]
\includegraphics[width=3.4in]{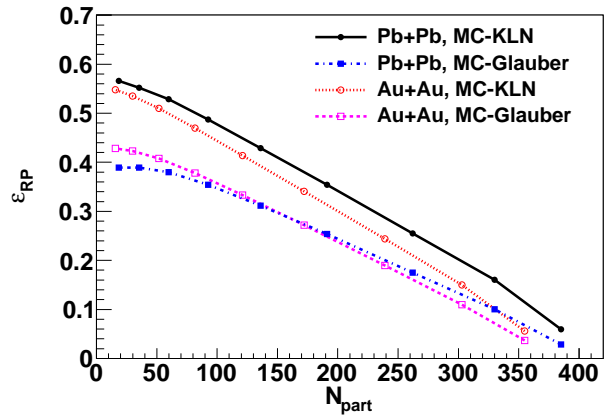}
\caption{(Color online)
Eccentricity with respect to the reaction plane as a function of 
$N_{\mathrm{part}}$
in Pb+Pb collisions at $\sqrt{s_{NN}}=$ 2.76 TeV and
in Au+Au collisions at $\sqrt{s_{NN}}=$ 200 GeV.
Each point 
from right to left corresponds to 0-5, 5-10, 10-20,
20-30, 30-40, 40-50, 50-60, 60-70, and 70-80\% centrality, respectively. 
}
\label{fig:eccnpart}
\end{figure}


In Fig.~\ref{fig:nchnpart},
we calculate $dN_{\mathrm{ch}}/d\eta/(N_{\mathrm{part}}/2)$
as a function of $N_{\mathrm{part}}$
for initial conditions from the MC-Glauber and the MC-KLN models
and compare them with data.
The experimental data point
in inelastic $p+p$ collisions at $\sqrt{s_{NN}}=2.36$ TeV \cite{Aamodt:2010ft}
is plotted 
at $N_{\mathrm{part}}=2$. The MC-KLN initialization
leads to remarkable agreement with 
the ALICE data.
On the other hand, it is difficult to fit
the data within the current two-component picture in the MC-Glauber model:
The results from the MC-Glauber initialization
with $\alpha = 0.08$ almost trace the ones from the MC-KLN
initialization and the ALICE data
except for 0-5\% and 70-80\% centrality.

\begin{figure*}[htb]
\includegraphics[width=3.4in]{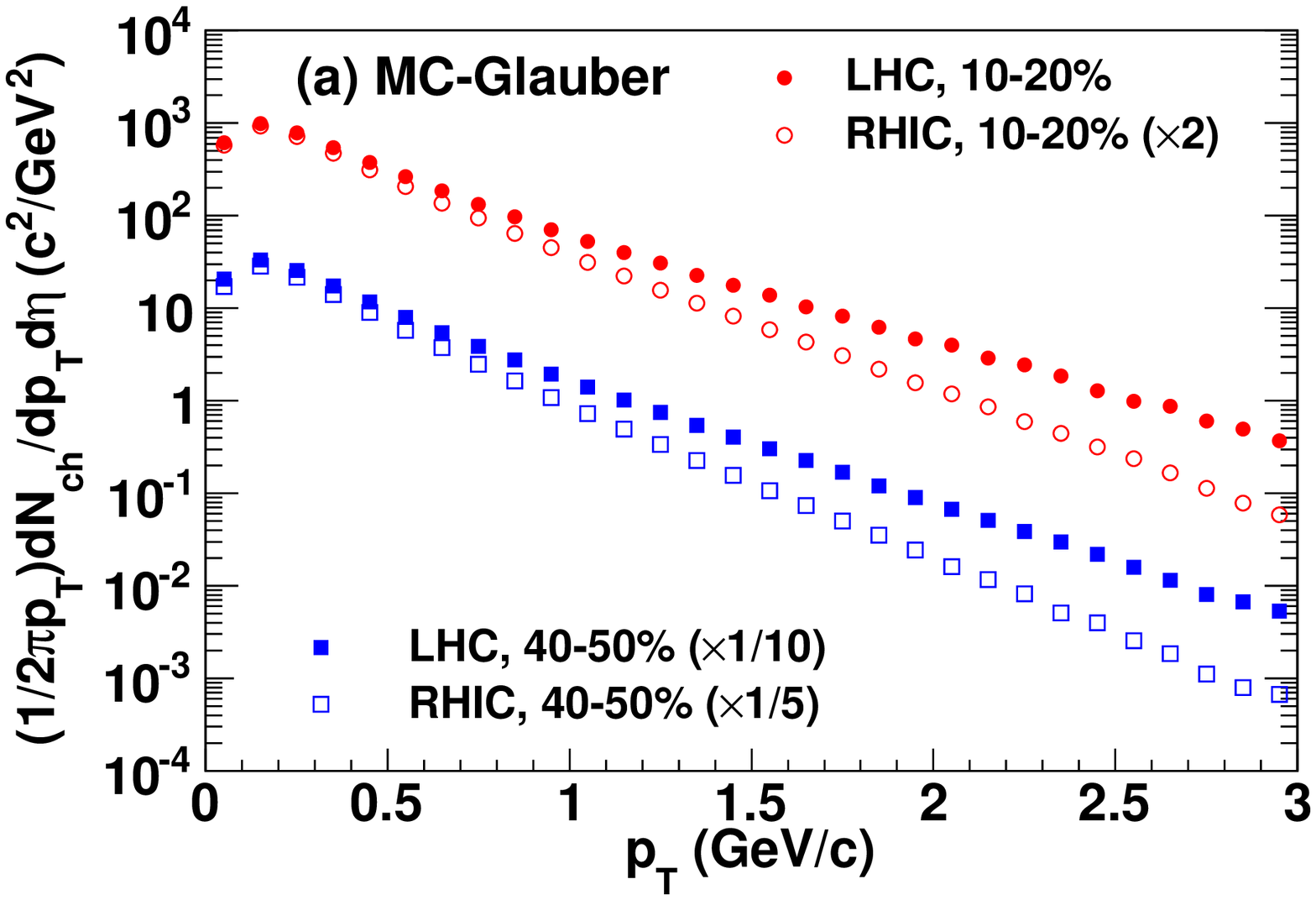}
\includegraphics[width=3.4in]{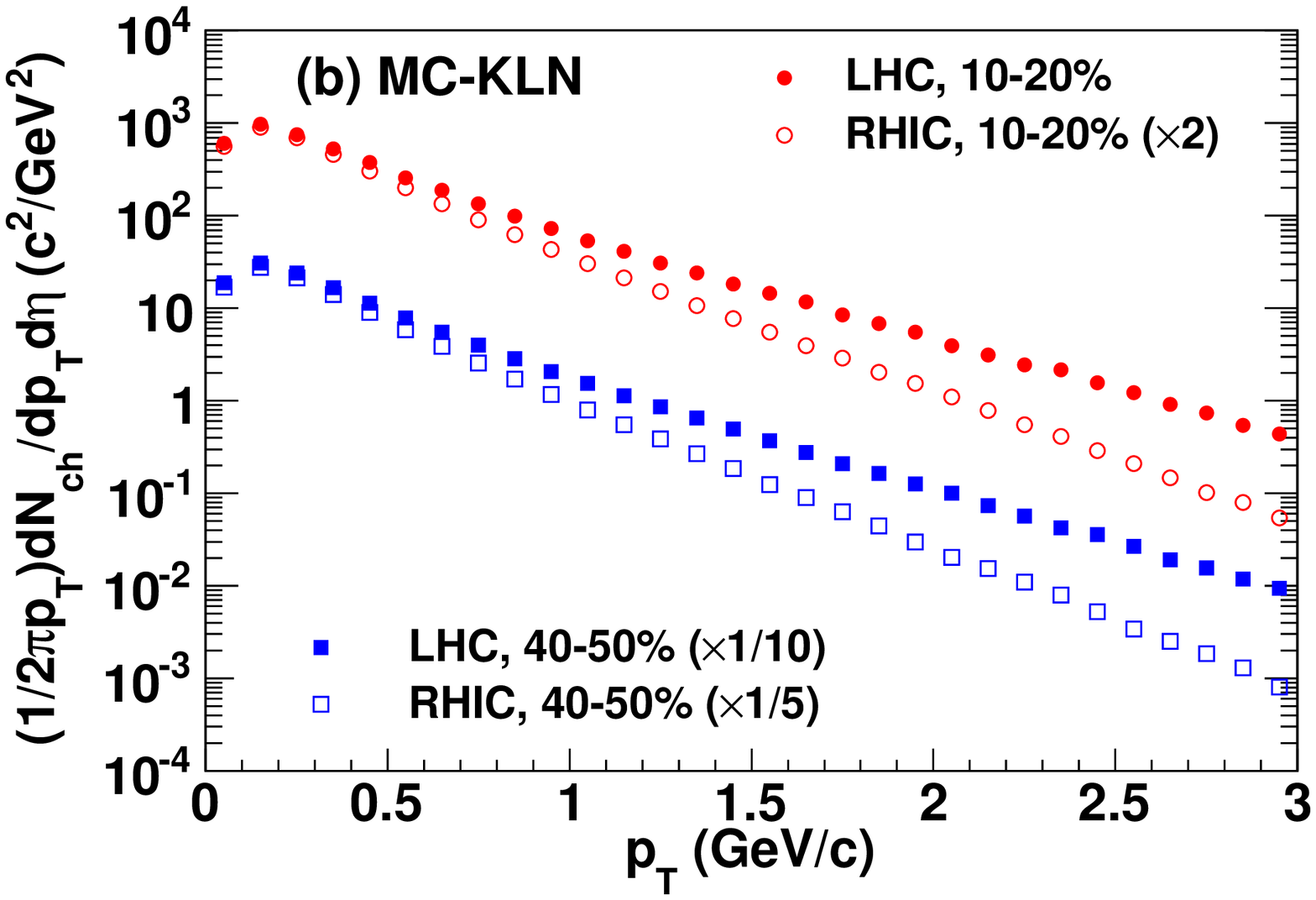}
\caption{(Color online)
Transverse momentum distribution of charged hadrons
at 10-20\% (circles) and 40-50\% (squares) centralities 
in Pb+Pb collisions at $\sqrt{s_{NN}}=$ 2.76 TeV (filled symbols)
and in Au+Au collisions at $\sqrt{s_{NN}}=$ 200 GeV (open symbols). 
Results from (a) the MC-Glauber initialization 
and (b) the MC-KLN initialization. 
For the sake of comparison and visibility, the spectra are scaled by
2, 1/10, and 1/5 for 10-20\% at RHIC, 40-50\% at LHC, and
40-50\% at RHIC, respectively.
}
\label{fig:dndpt}
\end{figure*}
%

Shown in Fig.~\ref{fig:eccnpart}
is the initial eccentricity 
with respect to reaction plane
as a function of $N_{\mathrm{part}}$
in Pb+Pb collisions at $\sqrt{s_{NN}}=$ 2.76 TeV
and in Au+Au collisions at $\sqrt{s_{NN}}=$ 200 GeV.
As previously known, the $k_t$-factorized formula of KLN model generates
larger eccentricity than the Glauber model does
\cite{Hirano:2005xf,fKLN}.
In the MC-KLN model,
eccentricity in Pb+Pb collisions
at $\sqrt{s_{NN}} = 2.76$ TeV
is slightly larger than that in Au+Au collisions
at $\sqrt{s_{NN}} = 200$ GeV 
when the centrality is fixed \cite{Hirano:2010jg}.
On the other hand, in the MC-Glauber model,
eccentricity in Pb+Pb collisions
at $\sqrt{s_{NN}} = 2.76$ TeV
is slightly smaller than
that in Au+Au collisions
at $\sqrt{s_{NN}} = 200$ GeV 
for a fixed centrality.

This is due to the smearing process we use to obtain a smooth
initial profile for hydrodynamic evolution. As mentioned, we use the
inelastic cross section in $p+p$ collisions, $\sigma_{\mathrm{in}}$,
to smear the distribution of collision points. This cross section 
is $\sim 1.5$ times larger at LHC than at RHIC, and thus the smearing
area, $S = \sigma_{\mathrm{in}}$ \cite{MCKLN}, is also larger at
LHC, and the eccentricity is reduced. Our smearing procedure also
leads to a smaller eccentricity than the conventional value of
MC-Glauber model\footnote{In the MC-Glauber model in the
literature~\cite{Miller:2007ri}, one assumes $\delta$ function
profile for each collision point in $\rho_{\mathrm{part}}$
distribution rather
than a box-like profile in the present work.}. The effect of
smearing is smaller in the MC-KLN initialization, and we have
checked that the eccentricity at LHC turns out to be essentially the
same as at RHIC when the smearing area is the same. Systematic
studies of initialization and its effects will be shown in a later
publication~\cite{HHN}.

Figure \ref{fig:dndpt} shows comparison of
transverse momentum distributions of charged hadrons
between RHIC and LHC energies at 10-20\% and 40-50\% centralities.
As clearly seen from figures,
the slope of $p_{T}$ spectra becomes flatter as collision energy
and, consequently, pressure of produced matter increases.
To quantify this,
we calculate mean $p_{T}$ of charged hadrons.
In the MC-Glauber initialization, 
mean $p_{T}$ increases from RHIC to LHC
by 21\% and 19\% in 10-20\% and 40-50\% centrality, respectively.
On the other hand, 
the corresponding fractions are  
25\% and 24\% in the MC-KLN initialization.
Since our calculations at RHIC were tuned to reproduce the
$p_T$-spectra, this means that at LHC the spectra calculated using
the MC-KLN initialization are slightly flatter than those calculated using
the MC-Glauber initialization.

\begin{figure*}[htb]
\includegraphics[width=3.4in]{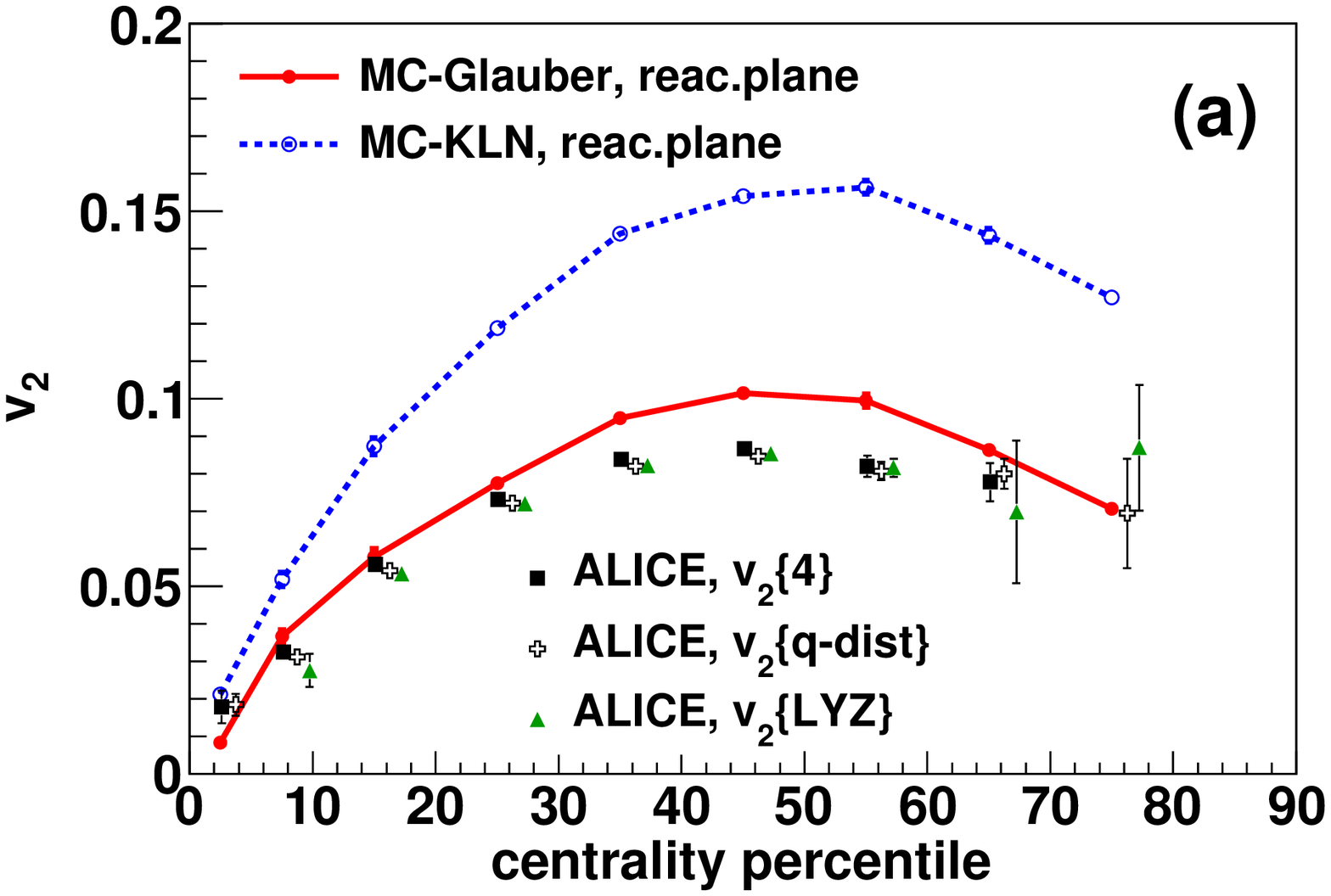}
\includegraphics[width=3.4in]{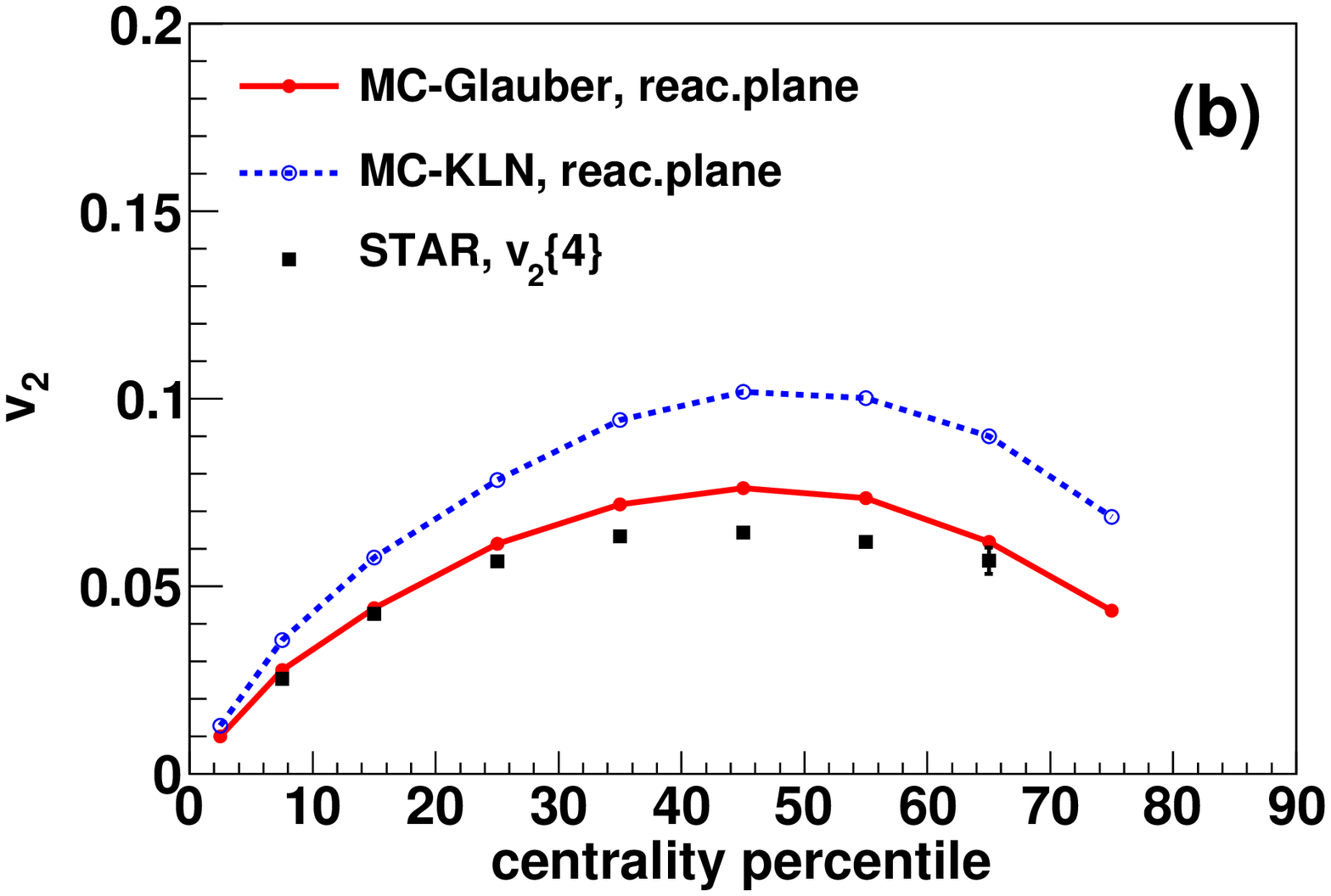}
\caption{(Color online)
Centrality dependences of $v_2$ 
for charged hadrons
with respect to reaction plane
(a) in Pb+Pb collisions
at $\sqrt{s_{NN}}=$ 2.76 TeV 
($\mid \eta \mid < 0.8$, $0.2 < p_{T} < 5$ GeV/$c$)
and (b) in Au+Au collisions at $\sqrt{s_{NN}}=$ 200 GeV 
($\mid \eta \mid < 1.3$, $0.15 < p_{T} < 2$ GeV/$c$)
are compared with ALICE \cite{Aamodt:2010pa} and STAR \cite{Adams:2004bi} 
$v_{2}$ data, respectively.
ALICE data points are shifted horizontally for visibility.
}
 \label{fig:v2cent}
 \end{figure*}
%

We compare integrated $v_2$ for charged hadrons with respect to
reaction plane with the ALICE \cite{Aamodt:2010pa} and STAR
\cite{Adams:2004bi} $v_{2}\{4\}$ data in Fig.~\ref{fig:v2cent}.
When evaluating the integrated $v_2$, we take account of both
transverse momentum and pseudorapidity acceptance as done in the
experiments, i.e. $0.2 < p_{T} < 5.0$ GeV/$c$ and 
$\mid \eta \mid < 0.8$ for ALICE, and $0.15 < p_{T} < 2.0$ GeV/$c$
and $\mid \eta \mid < 1.3$ for STAR. We want to emphasize that not
only the $p_T$ cut~\cite{Luzum:2010ag}, but also the pseudorapidity
cut plays an important role in a consistent comparison with the
data. Due to the Jacobian for the change of variables from rapidity
$y$ to pseudorapidity $\eta$, $v_2(y=0) < v_2(\eta=0)$ for positive
elliptic flow \cite{Kolb:2001yi}\footnote{Notice that even if one
assumes the Bjorken scaling solution, one has to consider the
pseudorapidity acceptance since $v_{2}(\eta)$ is not constant even
if $v_2(y)$ is~\cite{Kolb:2001yi}.}. In the case of the
MC-Glauber (MC-KLN) initialization in 40-50\% centrality, $v_{2}$
integrated over the whole $p_{T}$ region is $\sim$14\%
($\sim$10\%) larger at $\eta=0$ than at $y=0$.

When the MC-Glauber model is employed
for initial profiles,
centrality dependence of integrated $v_2$
from the hybrid approach
almost agrees with both ALICE and STAR data.
Since eccentricity fluctuation contributes little and negatively
to $v_{2}\{4\}$ in non-Gaussian distribution of
eccentricity fluctuation \cite{Voloshin:2007pc,Ollitrault:2009ie},
this indicates there is only
little room for the QGP viscosity in the model calculation.
On the other hand, apparent discrepancy
between the results from the MC-KLN initialization
and the ALICE and STAR data
means that 
viscous corrections during the hydrodynamic evolution are required.

From RHIC to LHC, the $p_{T}$-integrated $v_2(\mid \eta \mid < 0.8)$
increases by 24\% and 25\% in 10-20\% and 40-50\% centrality, respectively,
in the MC-Glauber initialization.
On the other hand, in the MC-KLN initialization, the increase
reaches 42\% and 44\% 
 in 10-20\% and 40-50\% centrality, respectively.
Since eccentricity does not change significantly (at most $\pm 6$\%
in 40-50\% centrality)
from RHIC to LHC as shown in Fig.~\ref{fig:eccnpart},
the significant increase of integrated $v_{2}$
must be attributed to a change in transverse 
dynamics.

%
\begin{figure*}[htb]
\includegraphics[width=3.4in]{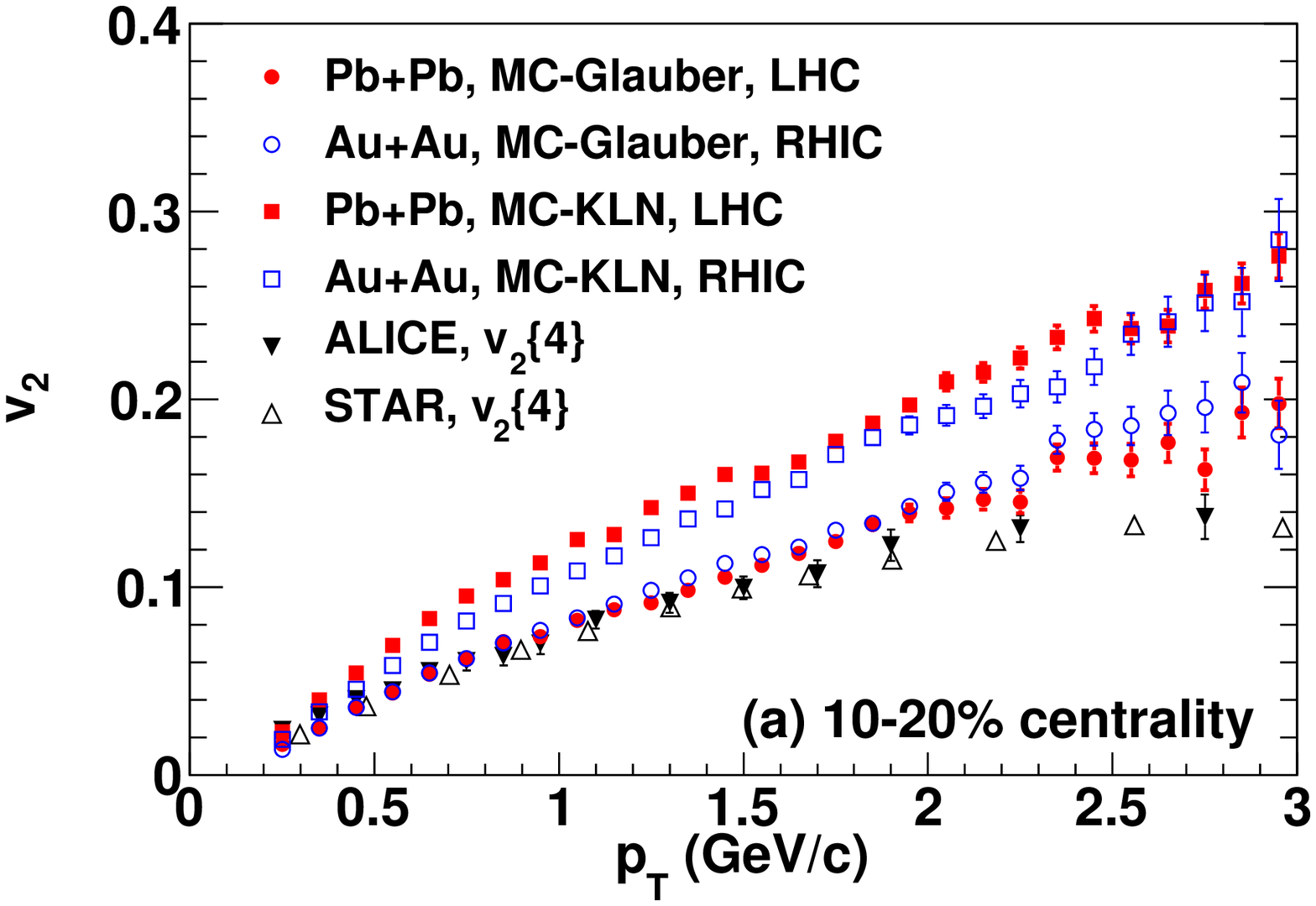}
\includegraphics[width=3.4in]{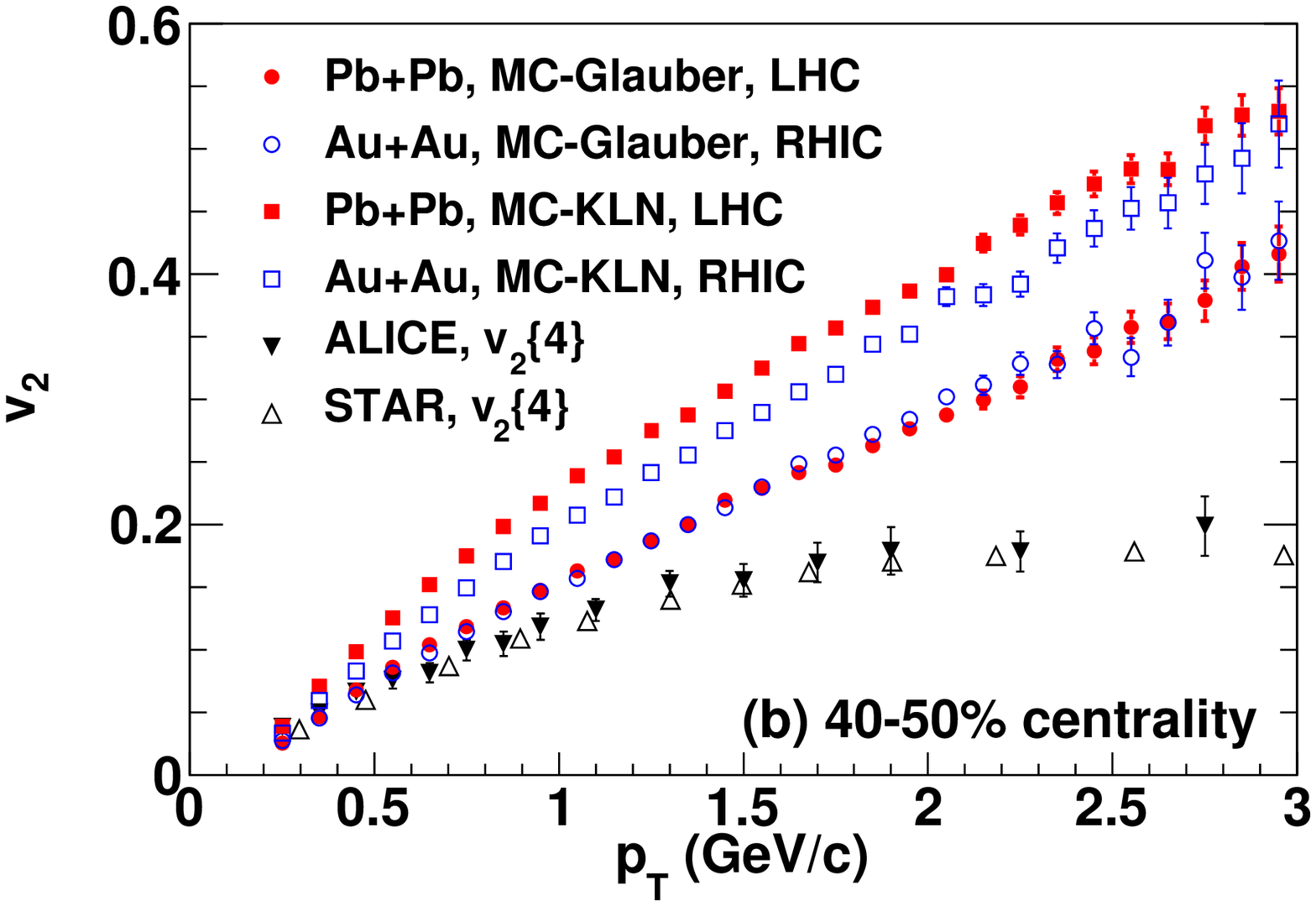}
\caption{(Color online)
Transverse momentum dependences of $v_2$ 
for charged hadrons
in the MC-Glauber (circles) and the MC-KLN (squares) initialization
are compared with ALICE \cite{Aamodt:2010pa} 
(triangles) and STAR \cite{Adams:2004bi} (band) $v_{2}\{4\}$ data
in (a) 10-20\% centrality and (b) 40-50\% centrality.
We take account of
pseudorapidity cut, $\mid \eta \mid < 0.8$ (1.3), in 
the ALICE (STAR) data.
}
 \label{fig:v2pt}
 \end{figure*}
%

Finally, we compare $v_{2}(p_{T})$ of charged hadrons
with ALICE \cite{Aamodt:2010pa} and STAR \cite{Adams:2004bi} data in
10-20\% (Fig.~\ref{fig:v2pt}(a)) and 40-50\% (Fig.~\ref{fig:v2pt}(b))
centrality.  Interestingly, the data at LHC agrees with the data
at RHIC within errors. The calculated $v_2(p_T)$ shows similar
independence of collision energy when MC-Glauber initialization is
used, whereas MC-KLN initialization leads to a slightly larger
$v_2(p_T)$ at the larger energy. For MC-Glauber results, the fit to
data is fair below $p_{T} \sim$ 1.5 GeV/$c$ and $p_{T} \sim$ 0.8
GeV/$c$ momenta in the 10-20\% and 40-50\% centralities,
respectively. Results from the MC-KLN initialization at both
energies are significantly larger than experimental data in the
whole $p_{T}$ region, which again indicates necessity of viscous
corrections in hydrodynamic evolution. For both initializations the
difference between the data and the calculated $v_2(p_T)$ is larger
in more peripheral collisions. This too can be understood as an
indication of viscosity, since the more peripheral the collision,
the smaller the system and the more anisotropic its shape, and both
of these qualities enhance the dissipative effects.

Due to the relationships among the $p_T$ spectrum, $p_T$
averaged $v_2$, and $p_T$ differential $v_2(p_T)$, the flatter the
$p_T$ spectrum, the larger the $v_2$ even if $v_2(p_T)$ stays the
same. It is also worth noticing that the steeper the slope of
$v_2(p_T)$, the larger the increase in $v_2$ for the same increase
in mean $p_T$. This is the main reason why quite a similar increase
of mean $p_T$ for both MC-Glauber and MC-KLN initializations leads
to much larger increase of $v_2$ for MC-KLN than for MC-Glauber
initialization.


At the time of this writing, the initial state of the fluid
dynamical expansion of heavy-ion collisions at ultrarelativistic
energies is quite uncertain.
This has been a longstanding issue
in the physics of heavy ion collisions
which must be by all means resolved.
If color glass condensate (CGC) \cite{Gelis:2010nm}
initial conditions, like the ones
obtained using the MC-KLN model in the present work,
are realized in nature 
at both RHIC and LHC energies,
the larger deviation of $v_{2}$ from the data at LHC than at RHIC
in Figs.~\ref{fig:v2cent} and \ref{fig:v2pt}
could mean that viscous effects are larger at LHC than at RHIC.
This can indicate a larger specific shear viscosity, $\eta/s$, at
larger temperatures.
For a better interpretation of current experimental data, 
the issue should be clarified in near future by determining the
initial conditions better
and by a more detailed analysis 
using a hybrid model of viscous hydrodynamics and hadron cascade 
\cite{Song:2010mg}.

In summary, we calculated transverse momentum distribution
of charged hadrons, centrality dependence of integrated
elliptic flow parameter $v_{2}$,
and differential elliptic flow $v_{2}(p_{T})$
in Pb+Pb collisions at $\sqrt{s_{NN}}=$ 2.76 TeV
and in Au+Au collisions at $\sqrt{s_{NN}}=$ 200 GeV.
We compared $v_{2}$ and $v_{2}(p_{T})$
with respect to reaction plane
from the hybrid model
with $v_2$ data mainly obtained from the 4-particle cumulant method.
Transverse momentum distributions become
harder, whereas
the shape of $v_{2}(p_{T})$
does not change so much as the collision energy increases.
Thus the increase in $p_T$-integrated $v_2$ is due to the increase in
mean $p_T$.
However, the intrinsic slope of $v_2(p_{T})$
depends on the initialization: The slope
from the MC-KLN initialization is steeper
than that from the MC-Glauber initialization,
and thus essentially the same change of mean $p_T$ leads to
larger increase of $p_T$-integrated $v_2$ for MC-KLN initialization
than for MC-Glauber initialization. The larger difference between
the data and our MC-KLN result at LHC than at RHIC may indicate larger
dissipative effects at LHC than at RHIC. All this 
emphasizes the
importance of understanding
initial conditions in relativistic heavy ion collisions
towards extracting the bulk and transport properties
of the QGP. 
In future, it would be interesting to
compare our results with data obtained using a more sophisticated
elliptic flow analysis \cite{Ollitrault:2009ie},
in which both non-flow and eccentricity
fluctuation effects are removed.

\acknowledgments
We would like to thank R.~Snellings
for providing us with the experimental data.
The work of T.H. (Y.N.) was partly supported by
Grant-in-Aid for Scientific Research
No.~22740151 (No.~20540276).
T.H. is also supported under
Excellent Young Researchers Oversea Visit Program
(No.~21-3383)
by Japan Society for the Promotion of Science.
P.H.'s work is supported by the ExtreMe Matter Institute (EMMI)
and BMBF contract No.~06FY9092.
T.H. thanks X.~N.~Wang and H.~Song for fruitful discussion
and members in the nuclear theory group
at Lawrence Berkeley National Laboratory
for a kind hospitality during his sabbatical stay.



\begin{thebibliography}{99}


\bibitem{Ollitrault}
  J.~Y.~Ollitrault,
  Phys.\ Rev.\ D {\bf 46}, 229 (1992).

\bibitem{Gyulassy:2004vg}
  M.~Gyulassy,
  arXiv:nucl-th/0403032.

\bibitem{sQGP}
  T.~D.~Lee,
  Nucl.\ Phys.\  A {\bf 750}, 1 (2005);
  M.~Gyulassy and L.~McLerran,
  Nucl.\ Phys.\  A {\bf 750}, 30 (2005);
  E.~V.~Shuryak,
  Nucl.\ Phys.\  A {\bf 750}, 64 (2005).


\bibitem{BNL}\url{http://www.bnl.gov/bnlweb/pubaf/pr/PR_display.asp?prID=05-38}


\bibitem{Aamodt:2010pa}
  K.~Aamodt {\it et al.}  [The ALICE Collaboration],
  arXiv:1011.3914 [nucl-ex].

\bibitem{Hirano:2010jg}
  T.~Hirano, P.~Huovinen, and Y.~Nara,
  arXiv:1010.6222 [nucl-th].


\bibitem{Aamodt:2010pb}
  K.~Aamodt {\it et al.}  [The ALICE Collaboration],
  arXiv:1011.3916 [nucl-ex].


\bibitem{Hirano:2001eu}
  T.~Hirano,
  Phys.\ Rev.\  C {\bf 65}, 011901 (2002);
  T.~Hirano and K.~Tsuda,
  Phys.\ Rev.\  C {\bf 66}, 054905 (2002).




\bibitem{JAM}
Y.~Nara, N.~Otuka, A.~Ohnishi, K.~Niita, and S.~Chiba,
 Phys.\ Rev.\  C {\bf 61}, 024901 (2000);
\url{http://quark.phy.bnl.gov/~ynara/jam/}







\bibitem{Cheng:2007jq}
  M.~Cheng {\it et al.},
  Phys.\ Rev.\  D {\bf 77}, 014511 (2008).

\bibitem{Bazavov:2009zn}
  A.~Bazavov {\it et al.},
  Phys.\ Rev.\  D {\bf 80}, 014504 (2009).


\bibitem{Huovinen:2009yb}
  P.~Huovinen and P.~Petreczky,
  Nucl.\ Phys.\  A {\bf 837}, 26 (2010).


\bibitem{EoSsite}
\url{https://wiki.bnl.gov/hhic/index.php/Lattice_calculatons_of_Equation_of_State}
and \url{https://wiki.bnl.gov/TECHQM/index.php/QCD_Equation_of_State}


\bibitem{Bjorken:1982qr}
  J.~D.~Bjorken,
  Phys.\ Rev.\  D {\bf 27}, 140 (1983).

\bibitem{Miller:2007ri}
  M.~L.~Miller, K.~Reygers, S.~J.~Sanders, and P.~Steinberg,
  Ann.\ Rev.\ Nucl.\ Part.\ Sci.\  {\bf 57}, 205 (2007).

\bibitem{MCKLN}
H. J. Drescher and Y. Nara, Phys. Rev. {\bf C 75}, 034905 (2007);
{\bf 76}, 041903(R) (2007);
\url{http://www.aiu.ac.jp/~ynara/mckln/}


\bibitem{Borghini:2000sa}
  N.~Borghini, P.~M.~Dinh, and J.~Y.~Ollitrault,
  Phys.\ Rev.\  C {\bf 63}, 054906 (2001);
  C {\bf 64}, 054901 (2001).


\bibitem{Voloshin:2007pc}
  S.~A.~Voloshin, A.~M.~Poskanzer, A.~Tang, and G.~Wang,
  Phys.\ Lett.\  B {\bf 659}, 537 (2008).

\bibitem{Ollitrault:2009ie}
  J.~Y.~Ollitrault, A.~M.~Poskanzer, and S.~A.~Voloshin,
  Phys.\ Rev.\  C {\bf 80}, 014904 (2009).


\bibitem{Hirano:2009ah}
  T.~Hirano and Y.~Nara,
  Phys.\ Rev.\  C {\bf 79}, 064904 (2009).


\bibitem{fKLN}
A. Adil, H. J. Drescher, A. Dumitru, A. Hayashigaki, 
and Y. Nara, Phys. Rev. C {\bf 74}, 044905 (2006).

\bibitem{KLN}
D. Kharzeev and M. Nardi, Phys. Lett. {\bf B507}, 121 (2001);
D. Kharzeev and E. Levin, ibid. {\bf B523}, 79 (2001); D. Kharzeev,
E. Levin, and M. Nardi, Phys. Rev. C {\bf 71}, 054903 (2005); Nucl.
Phys. {\bf A730}, 448 (2004).


\bibitem{Adler:2003cb}
  S.~S.~Adler {\it et al.}  [PHENIX Collaboration],
  Phys.\ Rev.\  C {\bf 69}, 034909 (2004).



\bibitem{ALICEdNdeta}
  K.~Aamodt {\it et al.}  [The ALICE Collaboration],
  arXiv:1012.1657 [nucl-ex].





\bibitem{Aamodt:2010ft}
  K.~Aamodt {\it et al.}  [ALICE Collaboration],
  Eur.\ Phys.\ J.\  C {\bf 68}, 89 (2010).


\bibitem{Hirano:2005xf}
  T.~Hirano, U.~W.~Heinz, D.~Kharzeev, R.~Lacey, and Y.~Nara,
  Phys.\ Lett.\  B {\bf 636}, 299 (2006).


\bibitem{HHN}
T.~Hirano, P.~Huovinen, and Y.~Nara, in preparation.

\bibitem{Adams:2004bi}
  J.~Adams {\it et al.}  [STAR Collaboration],
  Phys.\ Rev.\  C {\bf 72}, 014904 (2005).




\bibitem{Luzum:2010ag}
  M.~Luzum,
  arXiv:1011.5173 [nucl-th].

\bibitem{Kolb:2001yi}
  P.~F.~Kolb,
  Heavy Ion Phys.\  {\bf 15}, 279 (2002).

\bibitem{Gelis:2010nm}
  F.~Gelis, E.~Iancu, J.~Jalilian-Marian, and R.~Venugopalan,
  arXiv:1002.0333 [hep-ph].


\bibitem{Song:2010mg}
  H.~Song, S.~A.~Bass, U.~W.~Heinz, T.~Hirano, and C.~Shen,
  arXiv:1011.2783 [nucl-th].

\end{thebibliography}
\end{document}